# How Effective are COVID-19 Vaccine Health Messages in Reducing Vaccine Skepticism? Heterogeneity in Messages' Effectiveness by Just-World Beliefs

**Juliane Wiese**, corresponding author

Warwick Business School
University of Warwick
Scarman Road
Coventry CV4 7AL
United Kingdom

ORCID ID : 0000-0002-4314-5934

juliane.wiese@warwick.ac.uk
+33 6 68 88 18 27

**Nattavudh Powdthavee**

Nanyang Technological University
50 Nanyang Avenue
639798 Singapore

ORCID ID: 0000-0002-9345-4882

Declarations of interest: none.

## Abstract

To end the COVID-19 pandemic, policymakers have relied on various public health messages to boost vaccine take-up rates amongst people across wide political spectra, backgrounds, and worldviews. However, much less is understood about whether these messages affect different people in the same way. One source of heterogeneity is the belief in a just world (BJW), which is the belief that in general, good things happen to good people, and bad things happen to bad people. This study investigates the effectiveness of two common messages of the COVID-19 pandemic: vaccinate to protect yourself and vaccinate to protect others in your community. We then examine whether BJW moderates the effectiveness of these messages. We hypothesize that just-world believers react negatively to the prosocial pro-vaccine message, as it charges individuals with the responsibility to care for others around them. Using an unvaccinated sample of UK residents before vaccines were made widely available (N=526), we demonstrate that the individual-focused message significantly reduces overall vaccine skepticism, and that this effect is more robust for individuals with a low BJW, whereas the community-focused message does not. Our findings highlight the importance of individual differences in the reception of public health messages to reduce COVID-19 vaccine skepticism.

## 1. Introduction

Before the vaccine rollout in the UK, 28% of the British population, particularly those in Black and South Asian minority ethnic groups, were skeptical about getting vaccinated (Robertson *et al.*, 2021). To maximize vaccine take-up, governments have been delivering simple messages that emphasize people's responsibility to themselves *and* the community. For example, the National Health Services in the UK urges the public to "join the millions already vaccinated, to protect yourself and others" (NHS UK, 2021). These foci, given their central role in public health messaging during the COVID-19 pandemic so far, have shaped the two themes of messages examined in this study: individual and community responsibilities.

Despite the extensive literature on the framing approaches of public health messages around vaccines (*e.g.*, Gallagher & Updegraff, 2012; McPhee *et al.,* 2003; Kelly & Kornik, 2016), the overall effectiveness of COVID-19 vaccine messages on individual or community responsibility is currently imperfectly understood. While recent evidence suggests that individual-focused messages more effectively increase vaccine uptake and support for mandates than community-focused messages, these effects are heterogeneous across individualistic and communitarian worldviews (Yuan & Chu, 2022). Furthermore, we do not know which underlying beliefs about the vaccine are best addressed by these messages. Nevertheless, they continue to be used by public health officials worldwide.

In contexts of extreme urgency, who are the types of people who might respond poorly to these messages and experience stronger vaccine skepticism? We build our investigation around the strong theoretical link between belief in a just world (BJW) and vaccine skepticism. Just-world believers conceive a universal justice structure which holds that both normatively and positively speaking, good things tend to happen to good people and vice versa (Furnham, 2003). This adaptive function (Dalbert, 2009), manifesting at varying levels of intensity and therefore influencing a large portion of the population (White *et al.*, 2019), allows individuals to rationalize negative consequences in the world as justified, predictable, and manageable. Doing so promotes well-being and a sense of stability in the world (Correia et al., 2009; Jiang et al., 2016). In the context of the COVID-19 pandemic, where an unprecedented public health emergency and sweeping government regulations significantly reduced individual freedoms, just-world believers struggled to make sense of such undeserved restrictions. This sense of unfairness fosters a resistance against the government-promoted solution to the problem: specifically, a vaccine that has been developed in record speed. Suggestive evidence of this link between just-world believers and anti-vaxxers is demonstrated by their numerous shared psychological traits, including conspiracy thinking (Nestik *et al.,* 2020; Jolley & Douglas, 2014) and individualistic attitudes (Wenzel *et al*., 2017; Motta *et al.,* 2021). Government-sponsored pro-vaccine messages, particularly ones that focus on the responsibility we hold to our communities, are therefore likely to threaten the just-world believers' worldview, as their personal role in the pandemic is limited, and others' health outcomes are independent of their own decision to get vaccinated. Their worldview threatened, just-world believers defensively dismiss the message that threatens their BJW, and deny the existence of a problem in the first place (Furnham, 2003).

This study makes two main contributions to the literature. First, we experimentally investigate the effectiveness of two commonly used pro-vaccine messages. Second, we examine whether BJW moderates the effectiveness of each message. Given policymakers' priority to increase COVID-19 vaccine uptake, understanding individual differences in the messages' effectiveness by BJW is critical to understanding the potential threats to their overall effectiveness on the entire population.

## 2. Existing literature and hypotheses

Before the vaccine rollout, researchers' main concern was whether the COVID-19 vaccines safely reduce illness and transmissibility. Having established this (Katella, 2021; Pritchard *et al.*, 2021), vaccine uptake has emerged as a more enduring challenge for public health officials. A nationally representative survey of 316 Americans shows that demonstrating its efficacy and emphasizing the costs of the pandemic encourages vaccine uptake (Pogue *et al.*, 2020). However, their survey did not engage with messages that focus on the simple facts that give value to the vaccine: that it protects its recipients and their community. These facts have been central to policymakers' messaging during the COVID-19 pandemic, and there continues to be little empirical investigation into their effectiveness in shifting public perception around the vaccine's effectiveness.

The decision to vaccinate weighs the benefits against the risks of vaccination, which could range from fears of side effects and needles to mistrust of healthcare authorities. Previous research demonstrates the importance of highlighting vaccines' protective benefits, as doing so can crowd out concerns about risks (Porter *et al.,* 2018). Similarly, a COVID-19 vaccine message highlighting the vaccine's protective benefits to the individual has been shown to increase intended vaccine uptake (Yuan & Chu, 2022). Our work examines how such an individualistic message can drive the underlying beliefs around the vaccine's protective function to its recipients.

In addition, researchers have found prosocial vaccine messages to have a positive impact on vaccination rates (Betsch *et al.*, 2017; Betsch & Böhm, 2018; McPhee *et al.*, 2003). For example, messages that emphasize the benefits of an avian flu vaccine to others significantly increase vaccination intentions, compared to messages which emphasize its benefits to the individual (Kelly & Hornick, 2016). While these findings link the community-oriented message to increased vaccination intentions, they do not examine how such a message impacts beliefs around transmission rates, which is the mechanism that connects the prosocial messages with increased vaccine uptake. We aim to show experimentally that prosocial messages increase confidence in the underlying belief that the vaccines reduce transmission.

Based on this evidence, we predict that the individual message will more effectively decrease overall skepticism than the community message, and that this effect is driven by the fact that the individual message shifts the underlying belief that the vaccine protects its recipients. The prosocial messages will more moderately increase confidence that the vaccine reduces transmission.

Despite the predicted overall success of the two messages, the question remains around heterogeneous effects, specifically around moral worldviews that play a role in the decision to vaccinate. While Devereux *et al.* (2021) discover a link between stronger BJW and a greater likelihood to adhere to COVID-19 measures, such as social distancing, these measures come at essentially zero risk, resulting in a very different cost-benefit analysis. In contrast, accepting a vaccine requires accepting the risk of potential negative side-effects, and might therefore have a different relationship with BJW.

Demographic factors (Peretti-Watel *et al.*, 2020; Khubchandani *et al.*, 2021), psychological traits (Browne *et al.*, 2015; Jolley & Douglas, 2014), and beliefs about vaccine safety (Karlsson *et al.*, 2021) predict vaccine attitudes. However, studies that examine how such traits, like BJW, interfere with public health messages are scarce. While recent evidence has shown that people with more individualistic, rather than communitarian, values respond more favorably to individual-centered COVID-19 vaccine messages (Yuan & Chu, 2022), it remains unclear how such worldviews moderate individuals' understanding of the many ways in which the vaccine protects the public. Furthermore, rather than simply capturing individualistic or community-oriented worldviews, BJW contains a deeper moral around one's deservingness of one's place in the world, telling us more about the reasoning behind an individual's action (or inaction).

While people who see public health as a moral issue tend to consider prosocial (vs. self-centered) social distancing messages more persuasive (Luttrell & Petty, 2020), BJW is not an altruistic moral belief system. Instead, it holds individuals responsible for their own fate. BJW inherently commits fundamental attribution error, in which individuals place more weight on dispositional, as opposed to environmental or situational, factors (Ross, 1977). By further emphasising societal responsibility as a motive to get vaccinated, public health officials transfer the responsibility for a COVID patient's health onto the community's vaccination decision-making. This clashes with the tendency of just-world believers to blame patients for their own misfortunes and to separate the consequences of their own actions from the outcomes of others (Lerner & Simmons, 1966; Lucas *et al.*, 2009). Therefore, by asking people to take responsibility for others' health and safety during the COVID-19 pandemic, policymakers inevitably challenge the justice structure of the world in which individuals are responsible for their own fate. In response, just-world believers might discredit the vaccine altogether. We therefore hypothesise that for individuals with a strong BJW, the prosocial messages are less effective at reducing vaccine skepticism.

## 3. Method

### *3.1 Data*

In this pre-registered experiment (tinyurl.com/bxv23), 600 UK-based Prolific (www.prolific.co) users aged between 18 and 49 joined a longitudinal online study on attitudes towards COVID-19 and vaccination. At the time, the UK general public under 50 years of age was not yet eligible to receive a COVID-19 vaccine. Just over a quarter of the UK population had received its first dose, and only 1% of the population had received both doses (*Vaccinations in United Kingdom*, 30 April, 2021).

Part one of the study ($T_0$) took place on 24 February 2021, and part two ($T_1$) on 1 March 2021. We collected data at two points in time to reduce the likelihood that (i) participants suspect the study purpose and bias their responses, and (ii) participants' responses to vaccine skepticism questions are biased by exposure to questions around justice beliefs (Zizzo, 2010). Participants gave informed consent and were compensated £0.25 at $T_0$ and £1.00 at $T_1$.

527 participants (88%) remained at $T_1$ and were randomised evenly across Control, Individual-Treatment, and Prosocial-Treatment (N = 172, 181, and 174, respectively). Only one participant failed all three attention checks and was removed from the sample, resulting in 526 participants with complete longitudinal data. This sample size (i) allowed sufficient power for a reasonable minimal detectable effect size and (ii) is slightly larger than what was used in a similar research design studying BJW and climate change messaging (Feinberg & Willer, 2011). Of the final sample of 526 individuals, 70% were females, 87% were ethnically White, and 59% have an annual income of £30,000 or over. The mean age was 31. Balance checks confirm that our sample was balanced on observable characteristics across all groups; see Table A.1 in the appendix.

### *3.2 Measures and procedure*

#### *3.2.1 BJW scales*

Because vaccination evokes concepts of justice both for the individual and for society, participants completed the general BJW scale, six questions about the justice structure in the world in general (Dalbert *et al.*, 1987), and the personal BJW scale, seven questions which posit that the world is just for me personally but not for others (Dalbert, 1999) at $T_0$. The two scales have a correlation coefficient of 0.52. To attain a linear combination of BJW factors, we conducted a separate factor analysis on each scale, yielding two distinct factors ($\alpha = 0.78$ for general BJW and $\alpha = 0.88$ for personal BJW), and then conducted a factor analysis on these factors, resulting in a combined BJW factor ($\alpha = 0.68$); the factor analysis results are in *Table A.2*. The resulting combined BJW factor was standardized to a mean of 0 and a standard deviation of 1. It was transformed into a dummy variable which marks above- or below-median strength of BJW. This allows us to investigate the differential effects of the treatments on vaccine skepticism by the strength of BJW.

#### *3.2.2 Vaccine skepticism*

At $T_0$ and $T_1$, participants completed four questions on COVID-19 vaccine skepticism, with possible answers ranging from 0 (*not at all certain/likely*) to 100 (*extremely certain/likely)*. The precise wording of the questions was:

- "*How certain are you that the COVID-19 vaccines are a useful tool in fighting the pandemic?*"
- "*How likely are you to accept the COVID-19 vaccine when offered?*"
- "*How certain are you that the COVID-19 vaccine reduces transmission between individuals?*"
- "*How certain are you that the COVID-19 vaccine would prevent you personally from getting very ill due to COVID-19?*"

For simplicity, we reversed the responses so that higher values represent higher levels of vaccine skepticism in each of the four outcomes. The baseline mean responses are 16.8, 13.0, 29.1, and 22.4, respectively, which suggest that at $T_0$, the study population was relatively prepared to take the vaccine but was more skeptical of its illness and transmission prevention. These outcomes are moderately correlated, with correlations ranging from 0.53 to 0.76. To circumvent the multiple comparisons problem, we also derived an overall skepticism outcome by conducting a factor analysis on the four reversed individual skepticism variables for both outcomes at $T_0$ ($\alpha = 0.87$) and $T_1$ ($\alpha = 0.89$); see *Tables A.3 and A.4* in the appendix for the estimates. All skepticism variables were standardized to have a mean of 0 and a standard deviation of 1 and were included in analysis.

At $T_1$, 5 days after $T_0$, participants were randomised into one of three groups: control (no article), individual, and community responsibility treatment. In both treatments, participants were asked to read a news-style article. The articles, Figure A.3.1 in the appendix, are identical in the first paragraphs, which discuss the context of the pandemic and vaccine development at the time of writing. They deviate towards the end by treatment group. The individual responsibility article explains that the vaccine reduces the risk of severe COVID-19 illness to vaccine recipients, and the prosocial article explains that to combat the virus, individuals must accept the vaccine to reduce community transmission.

*3.2.3 Attention and manipulation check*

Participants in the treatment groups were asked two fact-based questions from the article, as well as whether taking the recommended steps during the pandemic will mainly protect them, or mainly protect others, from COVID-19 illness. Amongst the final sample of participants who passed all three attention checks, we find a significant difference between the two treatments on the manipulation-check item, $t(352) = 13.64$, $p = 0.000$ for indicating that the vaccine protects yourself, and $t(352) = -13.07$, $p = 0.000$ for indicating that the vaccine protects others.

*3.2.4 Sociodemographic controls*

Participants also completed a post-experiment questionnaire, which elicited their ethnicity, education level, region, income, political views, optimism, risk attitudes, COVID-19 history, and adhesion to government guidelines. Age and gender were collected automatically by Prolific.

Figure A.1 shows the procedural flow of the experiment and consort diagram, and Figures A.2 and A.3 present screenshots of the materials used.

### *3.3 Analysis*

We conduct all analyses of vaccine skepticism using Ordinary Least Squares (OLS) regression with robust standard errors clustered on the participant-level. Our primary analysis examines the treatment effects on the overall vaccine skepticism factor. We regress Equation (1) and present the results in column 3 of Table 1:

$$\Delta S_{it} = a + \beta_1 T_i + X_i'\gamma + \beta_2 BJW_i + \beta_3 (T_i \times BJW_i) + e, \quad (1)$$

where $i = 1, \ldots, N; t = 1, \ldots, 2$. $\Delta S_{it}$ represents the change in the overall vaccine skepticism factor from $t$=0 to $t$=1, where a higher value represents greater vaccine skepticism; $T_i$ represents the treatment condition (control, individual, or community message) and $\beta_1$ is the effect of this condition on skepticism; $X_i'$ represents the matrix of covariates, including a standardized optimism factor (α=0.8134), age, age-squared, gender dummy, £30,000+ annual income (vs. below £30,000 annual income) dummy, London (vs. non-London) dummy, undergraduate education (vs. non-undergraduate education) dummy, white (vs. non-white) dummy, Labour party (vs. non-Labour) dummy; $\beta_2$ is the effect of holding a strong (vs. weak) BJW; $\beta_3$ represents the interaction of treatment and BJW, *i.e.* the differential effect of the treatment when participants have either a stronger or a weaker BJW; and $e$ is the error term. Columns 1 and 2 model the parsimonious specifications of Eq. (1), with covariates excluding and including BJW, respectively.

Table 2 models the effects of the interaction between treatment and BJW on each of the four skepticism outcomes. Their forms are identical to Eq. (1), with the exception that the outcome variable is replaced by each of the four vaccine skepticism subscales, standardized to mean of 0 and standard deviation of 1.

$$\Delta\widehat{S_{jit}} = a + \beta_1 T_i + X_i'\gamma + \beta_2 BJW_i + \beta_3(T_i \times BJW_i) + e, \quad (2)$$

where $i = 1, \ldots, N; j = 1, \ldots, 4; t = 1, \ldots, 2$. Here, $\Delta\widehat{S_{ji}} = \widehat{S_{jit}} - \widehat{S_{jit-1}}$, where the notation $j$ represents different domains of beliefs, e.g., $S_{1i}$ represents the belief that the vaccine is not useful; $S_{2i}$ represents the likelihood of not accepting the vaccine; $S_{3i}$ represents the belief that the vaccine will not reduce transmission; and $S_{4i}$ represents the belief that the vaccine will not prevent serious illness. The rest of the specification is identical to Eq. (1).

Note that we deviate from the pre-registered document in two respects. First, we include an overall skepticism factor as an outcome variable in our primary analysis, circumventing the multiple comparisons problem in our primary analysis. Second, we run OLS regressions with standard errors clustered at the participant level as the primary analysis rather than using analysis of variance (ANOVA). This change is made due to the inclusion of continuous independent variables in the regression.

## 4. Results

### *4.1 Message effectiveness*

We begin by examining the within-person changes in vaccine skepticism by treatment group. As predicted, Figure 1 shows that the individual message significantly reduces overall skepticism by 0.04 standard deviation, compared to the control group which increases overall skepticism by 0.07 standard deviation (Wilcoxon signed-rank test, $p = 0.030$). There is weaker evidence that the community message also reduces overall skepticism, which decreased by 0.02 standard deviation (Wilcoxon signed-rank test, $p = 0.103$). Figure 1 thus provides raw data evidence that individual-focused public health message is most effective at reducing overall vaccine skepticism.

[Figure 1 here]

To understand this result more thoroughly, Table 1 estimates regression equations that adjust for other covariates, i.e., Eq.1. We find the regression results to be consistent with Figure 1's findings. The individual-focused message decreases overall skepticism more robustly than the community message, $\beta = -0.11$, [95% C.I.: -0.20, -0.02], $p = 0.014$, versus $\beta = -0.09$, [95% C.I.: - 0.19, 0.01], $p = 0.083$, respectively.

[Table 1 here]

***4.2 BJW as a moderator of pro-vaccine message impacts***

To formally test for the heterogeneous effect of public health messages by BJW, Tables 1 and 2 include the interaction terms between treatment and a high BJW dummy. Column 3 of Table 1 shows that for people with a low BJW, the individual message is extremely effective at lowering their overall skepticism factor, $\beta = - 0.19$, [95% C.I.: - 0.32, -0.06], $p = 0.004$. As discussed earlier, columns 1 and 2 of Table 1 demonstrate a greater effectiveness of the individual message on average. The results of column 3 suggest that the effectiveness of this individualistic message is more robust for people with a low BJW, whereas we see no such differential effect for the collective message. Figures 2 and 3 show this distinction visually, with the predictive margins plots of the control and individual treatment overlapping (Figure 2), and the predictive margins plots of the control and collective treatment (Figure 3) not overlapping. When examining the interaction regressions for each sub-scale of vaccine skepticism (Table 2), we find that the strong effect of the individual treatment on overall skepticism for people with a low BJW is driven by a reduction in skepticism around the belief that the vaccine will not prevent illness, $\beta = - 0.32$, [95% C.I.: - 0.50, -0.14], $p < 0.001$. This suggests that people with a low BJW, i.e. those who do **not** believe that there is a justice system which ensures that overall good things happen to good people and bad things happen to bad people, are extremely reactive to the individualistic message. It increases their confidence in the vaccine being able to protect them from serious illness. In other words, receiving the individualistic message, which accurately highlights that receiving the vaccine can prevent serious illness, correctly updates the beliefs around this issue for those with a low BJW, but not for those with a strong BJW. This suggests that for someone with a strong BJW, the belief in this just world order overpowers the belief in the science of the vaccine, as perhaps the deservingness of a person to fall ill would govern their likelihood of sickness moreso than the vaccine's protective properties.

[Figures 2 and 3 here]

Furthermore, we do not find evidence that people with a strong BJW react particularly poorly to the community message, $\beta = 0.03$, [95% C.I.: - 0.16, 0.22], $p = 0.778$. This suggests that a message which urges the public to take care of its community does not come into strong conflict with believers of a just world who may not feel responsible for the pandemic. This lack of resistance is consistent with just world believers' willingness to engage in other COVID-19 preventative measures (Devereux *et al.*, 2021).

[Table 2 here]

## 5. Discussions

Our findings that the individual and community messages concerning the COVID-19 vaccine can shift beliefs around the vaccine's various protective functions demonstrates an unsurprising link between the presentation of fact and its influence on a corresponding attitude. Nevertheless, in their desperate attempts to convince the public to get vaccinated, policymakers have sometimes turned to extreme measures, such as million-dollar lotteries, rifle giveaways, and free beer and donuts (Lewis 2021). However, while policymakers may have expected a clear increase in uptake, emerging evidence suggests that there is limited evidence in favor of these creative incentivizing strategies (Walkey *et al.,* 2022; Acharya & Dhakal, 2021), perhaps due to newfound suspicion of such gimmicky programs. Instead, policymakers should provide truthful information about the capacities of the COVID-19 vaccine, relying on existing evidence that these strategies effectively lower vaccine skepticism (Pennycook *et al*., 2020; Yuan & Chu, 2022).

Our messages do not easily shift the belief that the vaccine reduces transmission of the virus. This is especially important as new evidence emerges around the limited effectiveness of the vaccines against mutations of the coronavirus and in preventing transmission. Early studies suggest that the COVID-19 vaccines may not be as effective in preventing transmission as previously thought (Reuters, 2021). While policymakers should highlight the protective benefits of the vaccine, they must be cautious in not overstating the vaccine's effectiveness around transmission. Doing so could give vaccinated individuals a false sense of security, and ultimately reduced trust in public health authorities, resulting in less social distancing and respect for COVID-19 guidelines. As new scientific evidence about the vaccine emerges, officials must update their messaging content accordingly.

The literature shows that prosocial messages play an important role in motivating COVID-19 preventative actions, like signing up for contact-tracing apps (Jordan *et al.*, 2020). In contrast, vaccine skepticism responds differently. Consistent with previous findings (Yuan & Chu, 2022), we show that individual responsibility messages work as well, and sometimes better, than the community messages in reducing vaccine skepticism, depending on the dimension of skepticism in question. This discrepancy between non-vaccine COVID-19 prevention and vaccine messages could be because general preventative measures are perceived to be less risky than taking the vaccine. Riskier behaviors require more self-gain, which explains why the individual message is more successful.

Furthermore, the pro-vaccine messages used in this experiment affect different domains of vaccine skepticism differently. More specifically, they do not convince the population that the vaccine is useful to ending the pandemic, nor do they influence vaccination intentions. In the urgent pandemic context, while attitudes matter, vaccination behaviors are even more critical. Alternative strategies to motivate behavior must not be overlooked or confounded with strategies that target attitudes in future research.

When further examining heterogeneous treatment affects by intensity of BJW, we find that the overall success of the individual message is more robust among individuals with a low BJW, compared to those with a high BJW. The individual message, which focusses on the primary effect of the vaccine, may speak more particularly to people with a weak BJW because they see the world in a more factual, cartesian way. Someone with a strong BJW, on the other hand, may consider competing justice-related reasonings for the spread of or protection against COVID. The same is not true of the effects of the community message. Individuals with a strong BJW were found to be unmoved by the community message, possibly because this prosocial message sets an expectation that challenges the distribution of responsibility in a just world, as previously discussed. While individuals who see public health as a moral issue are more persuaded by other-focused (rather than self-focused) social distancing messages (Luttrell & Petty, 2020), BJW is not a worldview based on altruistic morals. Rather, where others may fall ill due to COVID-19, strong believers of a just world would blame the patients for their own misfortune, rather than assuming responsibility over the pandemic via mass collective vaccination.

Our results suggest that evidence-based messages (*e.g.:* the vaccine will protect you) have heterogeneous effects according to worldview. This heterogeneity replicates the findings of Yuan & Chu, who recently demonstrate that the individual-centered COVID-19 vaccine message is more impactful than a community-centered one, largely due to people whose worldview aligns with a more individualistic outlook (2022). Our studies differ in that we examine BJW, rather than individualism/communitarianism, and our sample was based in the UK, rather than the US. However, broadly speaking, the results confirm one another's findings, which is that the individual-centered message works best overall, but that this effect is driven largely by people with a worldview that places themselves, the individual, independent of a larger community or justice structure, at the center. Authorities ought to take into consideration the extent to which their vaccine messaging can have heterogeneous effects according to the worldviews of their population, especially as they encourage vaccine take-up amongst people with more extreme worldviews.

**6. Conclusions**

Simple messages that promote the COVID-19 vaccine effectively reduce vaccine skepticism of the corresponding beliefs around the vaccine's effectiveness. This reassuringly highlights the importance for policymakers to focus the information of their vaccination campaigns on the specific concerns of the public. The differences we find in effectiveness by psychological outlook are important for policymakers to consider, especially as the remaining unvaccinated likely hold more extreme world views. Messages that work well for people with low-level BJW evidently work less well for those with a more extreme worldview, suggesting that policymakers must reconsider how to motivate those harder-to-reach populations to get vaccinated. Custom messages that directly target people with such views could be an interesting line of research to follow.

This research is not without limitations. First, the data is restricted to a specific age-group in the United Kingdom and therefore has not been tested in other contexts, where just-world beliefs and vaccine skepticism differ. For example, in the United States, conservatism links with both BJW (Furnham, 2003) and COVID-19 vaccine skepticism (Latkin *et al.*, 2021), suggesting that BJW might be negatively correlated with pro-vaccine attitudes. Second, the sample in our study is not quota matched to the U.K. population, nor was it obtained using probability sampling. Hence, the results cannot be considered nationally representative, and there is likely a degree of selection bias amongst users of Prolific. Third, our dataset does not capture whether participants ultimately took up the vaccination, as it only captures attitudes and intentions. As previously discussed, behaviors in this context are more important than attitudes, and would be valuable to follow up on.

The authors declare no conflicts of interest.

**How Effective are COVID-19 Vaccine Health Messages in Reducing Vaccine Skepticism? Heterogeneity in Messages' Effectiveness by Just-World Beliefs**

Tables and Figures

**Figure 1.** Proportions of overall skepticism changes across control, individual, and community messages.

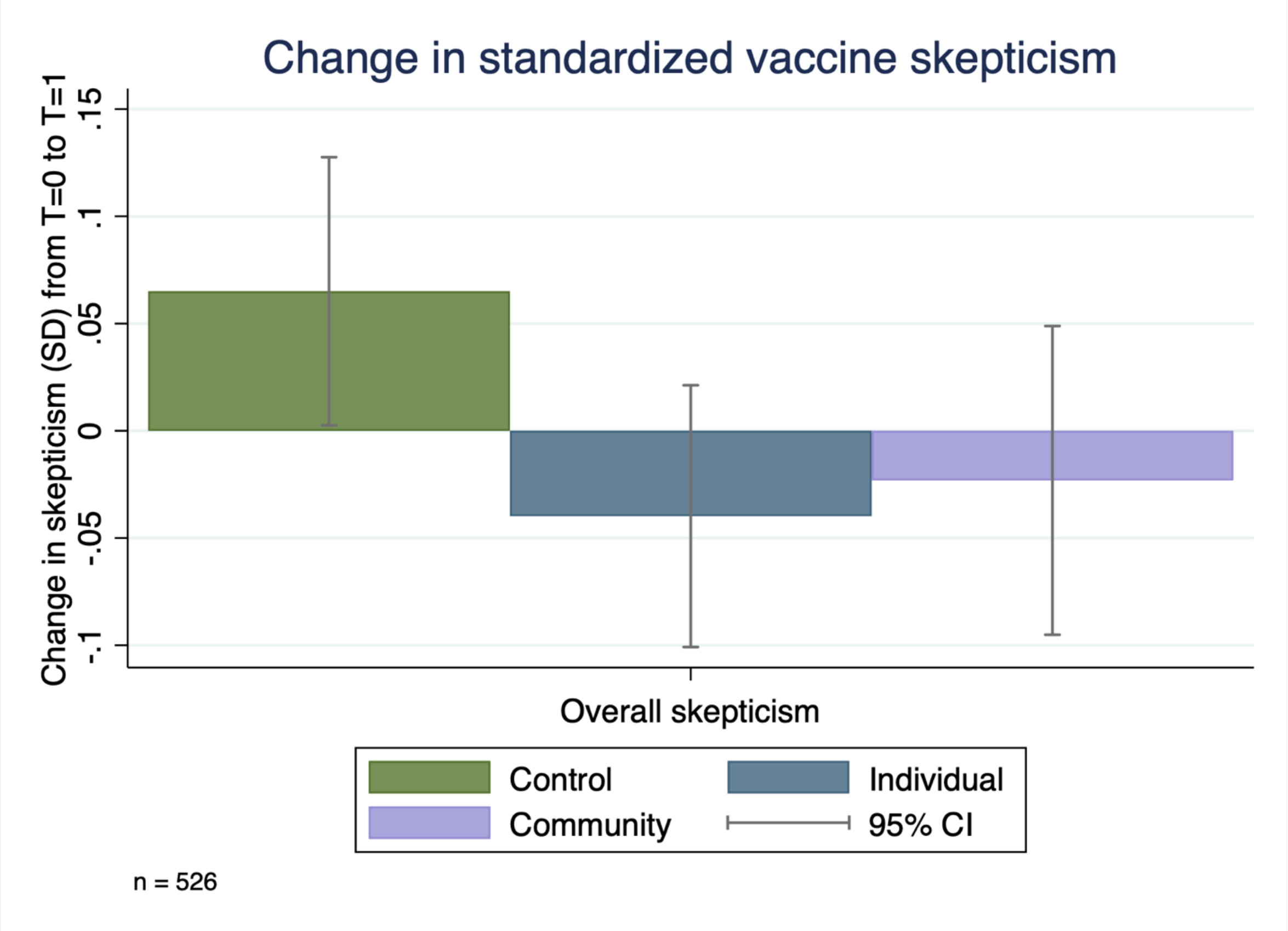

**Table 1: The effects of public health messages on overall vaccine skepticism factor outcome: OLS regressions**

| | (1)<br>Δ Skepticism factor (std) | (2)<br>Δ Skepticism factor (std) | (3)<br>Δ Skepticism factor (std) |
|---|---|---|---|
| Individual | -0.113** | -0.113** | -0.189*** |
| | (0.0453) | (0.0457) | (0.0657) |
| Community | -0.0872 | -0.0872 | -0.101 |
| | (0.0502) | (0.0502) | (0.0639) |
| High BJW | | 0.000134 | -0.0569 |
| | | (0.0434) | (0.0658) |
| Individual x High BJW | | | 0.147 |
| | | | (0.0911) |
| Community x High BJW | | | 0.0274 |
| | | | (0.0969) |
| Optimism (std) | -0.00429 | -0.00432 | -0.00569 |
| | (0.0201) | (0.0215) | (0.0215) |
| Age | 0.00640 | 0.00640 | 0.00625 |
| | (0.0190) | (0.0191) | (0.0188) |
| Age squared | -8.56e-05 | -8.56e-05 | -8.20e-05 |
| | (0.000293) | (0.000294) | (0.000290) |
| Female | -0.000917 | -0.000910 | 0.00305 |
| | (0.0438) | (0.0440) | (0.0436) |
| £ 30k+ | -0.0179 | -0.0179 | -0.0153 |
| | (0.0410) | (0.0414) | (0.0411) |
| London | -0.0313 | -0.0313 | -0.0292 |
| | (0.0635) | (0.0636) | (0.0633) |
| University+ | 0.0467 | 0.0467 | 0.0431 |
| | (0.0430) | (0.0434) | (0.0429) |
| White | 0.0210 | 0.0210 | 0.0254 |
| | (0.0703) | (0.0705) | (0.0708) |
| Labour | -0.00769 | -0.00768 | -0.00469 |
| | (0.0375) | (0.0372) | (0.0378) |
| Constant | -0.0700 | -0.0701 | -0.0499 |
| | (0.303) | (0.307) | (0.300) |
| Cluster individuals | 526 | 526 | 526 |
| R-squared | 0.024 | 0.024 | 0.029 |

**Note:** *** p<0.001, ** p<0.05. Robust standard errors clustered at the individual level and are in parentheses. Dependent variables represent the change from $T_0$ to $T_1$ and are standardized to have a mean of 0 and a standard deviation of 1.

**Table 2: The effects of public health messages on individual skepticism outcomes: OLS regressions with BJW interactions**

| | (1)<br>Δ Vaccine not useful (std) | (2)<br>Δ Not accept vaccine (std) | (3)<br>Δ Not reduce transmission (std) | (4)<br>Δ Not prevent illness (std) |
|---|---|---|---|---|
| Individual | -0.166 | -0.0212 | -0.0458 | -0.323*** |
| | (0.102) | (0.0466) | (0.107) | (0.0917) |
| Community | -0.101 | 0.0175 | -0.0904 | -0.139 |
| | (0.0951) | (0.0545) | (0.110) | (0.0972) |
| High BJW | -0.0605 | -0.00896 | 0.0244 | -0.103 |
| | (0.0951) | (0.0543) | (0.116) | (0.113) |
| Individual x High BJW | 0.147 | 0.0565 | 0.0416 | 0.214 |
| | (0.135) | (0.0790) | (0.158) | (0.146) |
| Community x High BJW | 0.0651 | -0.0288 | -0.189 | 0.0998 |
| | (0.134) | (0.0827) | (0.170) | (0.155) |
| Optimism (std) | -0.0281 | -4.72e-05 | 0.00949 | 0.0194 |
| | (0.0253) | (0.0188) | (0.0378) | (0.0344) |
| Age | -0.00199 | 0.0131 | 0.0512 | -0.00812 |
| | (0.0276) | (0.0156) | (0.0311) | (0.0281) |
| Age squared | 5.24e-05 | -0.000208 | -0.000845 | 0.000158 |
| | (0.000424) | (0.000227) | (0.000481) | (0.000433) |
| Female | 0.0213 | -0.0378 | 0.0621 | -0.0174 |
| | (0.0616) | (0.0397) | (0.0760) | (0.0696) |
| £ 30k+ | -0.00347 | 0.0365 | -0.0707 | -0.0745 |
| | (0.0606) | (0.0381) | (0.0706) | (0.0662) |
| London | -0.00687 | -0.0415 | 0.00830 | -0.0467 |
| | (0.0989) | (0.0448) | (0.0875) | (0.0791) |
| University+ | 0.0975 | -0.0117 | -0.0588 | 0.0427 |
| | (0.0601) | (0.0326) | (0.0678) | (0.0667) |
| White | 0.0705 | -0.0778 | 0.122 | -0.0191 |
| | (0.108) | (0.0591) | (0.118) | (0.115) |
| Labour | -0.0578 | 0.0123 | -0.0130 | 0.0311 |
| | (0.0557) | (0.0321) | (0.0687) | (0.0605) |
| Constant | -0.0239 | -0.114 | -0.716 | 0.284 |
| | (0.458) | (0.259) | (0.498) | (0.444) |
| Cluster individuals | 526 | 526 | 526 | 526 |
| R-squared | 0.023 | 0.021 | 0.032 | 0.034 |

**Note:** *** p<0.001, ** p<0.05. Robust standard errors clustered at the individual level and are in parentheses. Dependent variables represent the change from $T_0$ to $T_1$ and are standardized to have a mean of 0 and a standard deviation of 1.

**Figure 2: Predictive margins of the individual treatment and control group, over the standardized BJW factor**

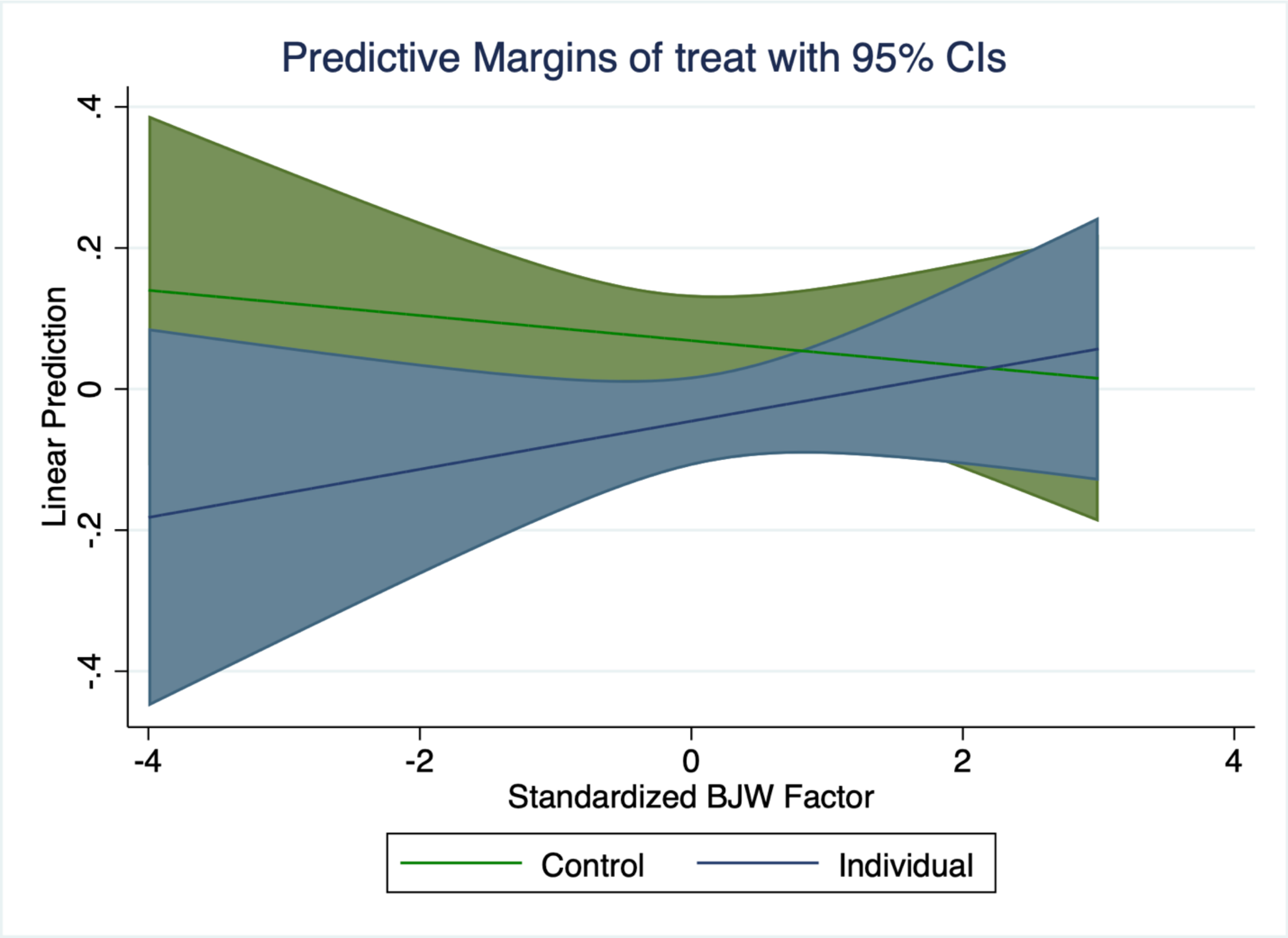

**Figure 3: Predictive margins of the community treatment and control group, over the standardized BJW factor**

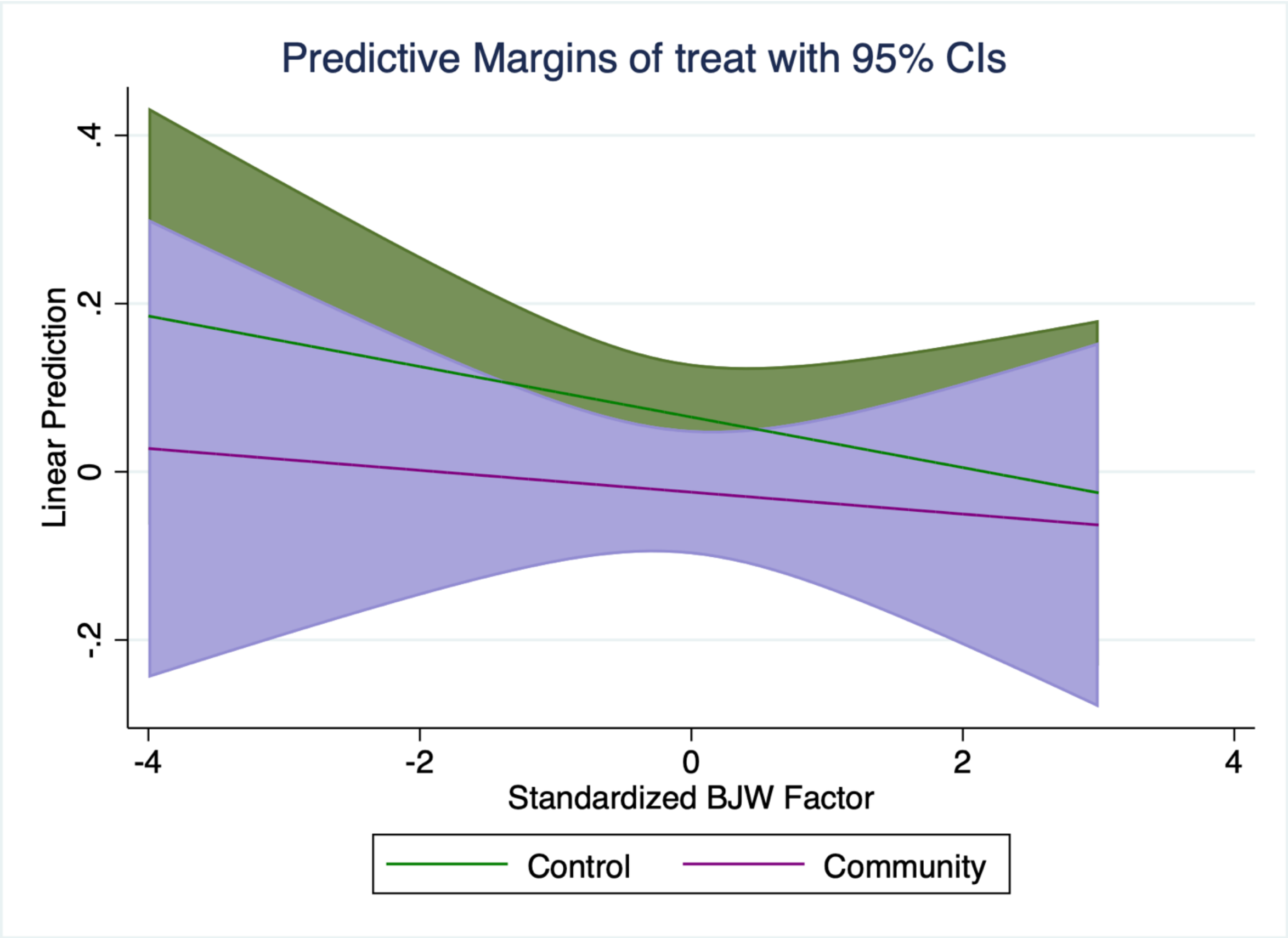

# How Effective are COVID-19 Vaccine Health Messages in Reducing Vaccine Skepticism? Heterogeneity in Messages' Effectiveness by Just-World Beliefs

Appendix

**Table A.1**: Balance checks on all observable characteristics amongst the final analysis sample.

| | | Control (0) | Individual (1) | Community (2) | (0) vs. (1), p-value | (0) vs. (2), p-value | (1) vs. (2), p-value |
|---|---|---|---|---|---|---|---|
| $T_0$(baseline) | Not vaccine useful | 17.0 | 16.2 | 17.3 | 0.704 | 0.904 | 0.616 |
| | | (1.5) | (1.4) | (1.6) | | | |
| | | *172* | *180* | *174* | | | |
| | Not accept vaccine | 12.5 | 12.6 | 13.9 | 0.988 | 0.628 | 0.625 |
| | | (1.9) | (1.7) | (2.0) | | | |
| | | *172* | *180* | *174* | | | |
| | Not reduce transmission | 29.9 | 28.7 | 29.0 | 0.645 | 0.670 | 0.983 |
| | | (1.9) | (1.9) | (2.0) | | | |
| | | *172* | *180* | *174* | | | |
| | Not prevent illness | 22.0 | 22.3 | 22.8 | 0.895 | 0.741 | 0.839 |
| | | (1.8) | (1.7) | (1.8) | | | |
| | | *172* | *180* | *174* | | | |
| $T_1$ (endline) | Not vaccine useful | 15.6 | 13.5 | 14.5 | 0.253 | 0.615 | 0.566 |
| | | (1.4) | (1.2) | (1.5) | | | |
| | | *172* | *180* | *174* | | | |
| | Not accept vaccine | 11.6 | 12.0 | 13.0 | 0.892 | 0.602 | 0.684 |
| | | (1.8) | (1.7) | (1.8) | | | |
| | | *172* | *180* | *174* | | | |
| | Not reduce transmission | 30.7 | 29.0 | 25.0 | 0.548 | 0.039 | 0.124 |
| | | (2.0) | (1.9) | (1.8) | | | |
| | | *172* | *180* | *174* | | | |
| | Not prevent illness | 22.1 | 17.8 | 20.8 | 0.054 | 0.581 | 0.187 |
| | | (1.7) | (1.5) | (1.7) | | | |
| | | *172* | *180* | *174* | | | |
| | Quartile 1 BJW | 0.3 | 0.2 | 0.3 | 0.626 | 0.671 | 0.358 |
| | | (0.0) | (0.0) | (0.0) | | | |
| | | *172* | *180* | *174* | | | |
| | Quartile 4 BJW | 0.2 | 0.3 | 0.3 | 0.529 | 0.352 | 0.756 |
| | | (0.0) | (0.0) | (0.0) | | | |
| | | *172* | *180* | *174* | | | |
| | Optimism factor (Std) | 0.0 | 0.0 | -0.0 | 0.928 | 0.779 | 0.850 |
| | | (0.1) | (0.1) | (0.1) | | | |
| | | 172 | 180 | 174 | | | |
| | Age | 30.4 | 31.0 | 31.5 | 0.479 | 0.231 | 0.600 |
| | | (0.7) | (0.6) | (0.7) | | | |
| | | *172* | *180* | *174* | | | |
| | Female | 0.7 | 0.8 | 0.7 | 0.309 | 0.459 | 0.781 |
| | | (0.0) | (0.0) | (0.0) | | | |
| | | *166* | *171* | *170* | | | |
| | £30,000+ | 0.7 | 0.7 | 0.8 | 0.987 | 0.455 | 0.423 |
| | | (0.1) | (0.0) | (0.1) | | | |
| | | *172* | *180* | *174* | | | |
| | London | 0.1 | 0.1 | 0.2 | 0.652 | 0.248 | 0.472 |
| | | (0.0) | (0.0) | (0.0) | | | |
| | | *172* | *180* | *174* | | | |
| | Undergraduate+ | 0.6 | 0.6 | 0.6 | 0.778 | 0.808 | 0.597 |
| | | (0.0) | (0.0) | (0.0) | | | |
| | | *171* | *179* | *174* | | | |
| | White | 0.9 | 0.9 | 0.9 | 0.525 | 0.724 | 0.779 |
| | | (0.0) | (0.0) | (0.0) | | | |
| | | *172* | *180* | *174* | | | |
| | Labour party | 0.3 | 0.4 | 0.4 | 0.825 | 0.438 | 0.578 |

| | | | |
|---|---|---|---|
| | (0.0) | (0.0) | (0.0) |
| | *169* | *172* | *172* |

**Note:** standard deviations in parenthesis, sample size of respondents in italics.

**Figure A.1**: Experimental process and consort diagram

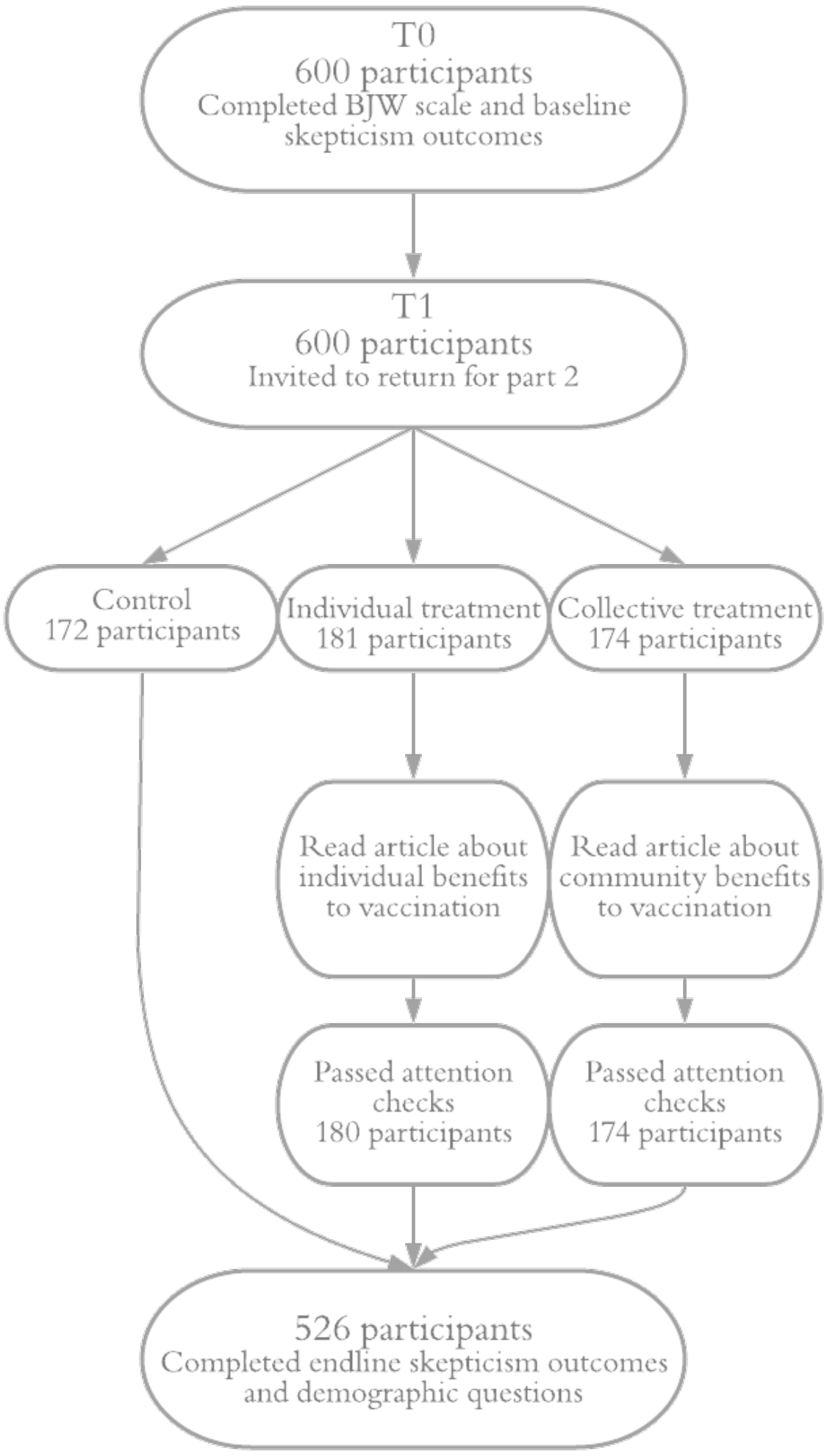

**Figure A.2:** Survey design: questions at $T_0$.

Please read each statement carefully and indicate the extent to which you personally agree or disagree with it.

| | Very strongly disagree | Disagree | Slightly disagree | Slightly agree | Agree | Very strongly agree |
|---|---|---|---|---|---|---|
| I think basically the world is a just place. | ○ | ○ | ○ | ○ | ○ | ○ |
| I believe that, by and large, people get what they deserve. | ○ | ○ | ○ | ○ | ○ | ○ |
| I am confident that justice always prevails over injustice. | ○ | ○ | ○ | ○ | ○ | ○ |
| I am convinced that in the long run, people will be compensated for injustices. | ○ | ○ | ○ | ○ | ○ | ○ |
| I firmly believe that injustices in all areas of life (e.g. professional, family, politics) are the exception rather than the rule. | ○ | ○ | ○ | ○ | ○ | ○ |
| I think people try to be fair when making important decisions. | ○ | ○ | ○ | ○ | ○ | ○ |

Please read each statement carefully and indicate the extent to which you personally agree or disagree with it.

| | Very strongly disagree | Disagree | Slightly disagree | Slightly agree | Agree | Very strongly agree |
|---|---|---|---|---|---|---|
| I believe that, by and large, I deserve what happens to me. | ○ | ○ | ○ | ○ | ○ | ○ |
| I am usually treated fairly. | ○ | ○ | ○ | ○ | ○ | ○ |
| I believe that I usually get what I deserve. | ○ | ○ | ○ | ○ | ○ | ○ |
| Overall, events in my life are just. | ○ | ○ | ○ | ○ | ○ | ○ |
| In my life injustice is the exception rather than the rule. | ○ | ○ | ○ | ○ | ○ | ○ |
| I believe that most of the things that happen in my life are fair. | ○ | ○ | ○ | ○ | ○ | ○ |
| I think that important decisions that are made concerning me are usually just. | ○ | ○ | ○ | ○ | ○ | ○ |

How certain are you that the COVID-19 vaccines are a useful tool in fighting the pandemic?

Not at all certain — Extremely certain

0 10 20 30 40 50 60 70 80 90 100

.

---

How likely are you to accept the COVID-19 vaccine when offered?

Not at all likely — Extremely likely

0 10 20 30 40 50 60 70 80 90 100

.

---

How certain are you that the COVID-19 vaccine reduces transmission between individuals?

0 10 20 30 40 50 60 70 80 90 100

.

---

How certain are you that the COVID-19 vaccine would prevent you personally from getting very ill due to COVID-19?

0 10 20 30 40 50 60 70 80 90 100

.

**Figure A.3.1:** Survey design: treatment messages at $T_1$. Control participants were asked to respond to the same four skepticism outcomes shown in Figure A.2. Individual (left) and community (right) messages participants were first asked to read the following fictitious news articles and were then prompted to respond to the four skepticism outcomes.

Below is a news story similar to other news stories you might have read before. Please read the story and respond to the questions that follow.

BOSTON -- “Coronavirus disease (COVID-19) is a highly contagious illness, caused by the transmission of the SARS-CoV-2 virus. First identified in December 2019, the virus has caused a pandemic that resulted in shutdowns all around the globe. It was first widely spread between humans at a wholesale seafood market in Wuhan, China. “This disease has wreaked havoc on the globe. Claiming millions of lives, this pandemic has created a clear demarcation in time: pre-covid, and post-covid,” says Professor Arthur Michali, a public health expert from a leading research university. “Before this pandemic, I could have attended a conference in Tokyo one day, led a research collaboration in Geneva the next, and arrived back in Boston the third day. This kind of travel is simply no longer possible under current circumstances, and it’s likely that this sort of behaviour contributed to the rapid spread of the disease worldwide.”

Professor Michali, who has won numerous awards for his research over the last two decades, is part of the COVID-19 Emergency Committee at the World Health Organisation. Amongst other topics, this committee is working to better understand the various responses and interventions that can help curb the spread of the disease.

Michali is co-authoring a forthcoming pamphlet, entitled “The COVID-19 Vaccine: what you need to know,” which aims to educate the average citizen about the currently circulating and forthcoming vaccines. The pamphlet describes COVID-19 as “a dangerous disease, particularly for the elderly and clinically vulnerable, as they are more likely to suffer severe, and possibly fatal, respiratory illness. Nevertheless, anyone, regardless of age or medical background, is at risk of suffering a harsh illness.” Michali wishes to emphasise that **the best thing you can do to protect yourself from this disease is to take up the vaccine when you are offered it**. “Some of the vaccines on the market are boasting 95% efficacy rates. This means that receiving the vaccine dramatically reduces your risk of developing serious COVID-19 symptoms if you are exposed to the virus later down the line.” Although experts are continuing to emphasise the importance of social distancing and wearing masks, these measures are not perfect, and there remains a risk of inadvertently catching the disease that could leave you bed-ridden for weeks, even months. Receiving the vaccine is the single most important step an individual can take to protect him or herself from the virus. Michali reflects in the concluding thoughts of the pamphlet, **“there is not much that we can control in times like these, but you need to do what you can to protect yourself in these uncertain times. Taking up the vaccine when offered is the best action you can take to keep yourself safe!”** Importantly, Michali wants individuals to remember that it is their personal responsibility to keep themselves protected.

Below is a news story similar to other news stories you might have read before. Please read the story and respond to the questions that follow.

BOSTON -- “Coronavirus disease (COVID-19) is a highly contagious illness, caused by the transmission of the SARS-CoV-2 virus. First identified in December 2019, the virus has caused a pandemic that resulted in shutdowns all around the globe. It was first widely spread between humans at a wholesale seafood market in Wuhan, China. “This disease has wreaked havoc on the globe. Claiming millions of lives, this pandemic has created a clear demarcation in time: pre-covid, and post-covid,” says Professor Arthur Michali, a public health expert from a leading research university. “Before this pandemic, I could have attended a conference in Tokyo one day, led a research collaboration in Geneva the next, and arrived back in Boston the third day. This kind of travel is simply no longer possible under current circumstances, and it’s likely that this sort of behaviour contributed to the rapid spread of the disease worldwide.”

Professor Michali, who has won numerous awards for his research over the last two decades, is part of the COVID-19 Emergency Committee at the World Health Organisation. Amongst other topics, this committee is working to better understand the various responses and interventions that can help curb the spread of the disease.

Michali is co-authoring a forthcoming pamphlet, entitled “The COVID-19 Vaccine: what you need to know,” which aims to educate the average citizen about the currently circulating and forthcoming vaccines. The pamphlet describes COVID-19 as “a dangerous disease, particularly for the elderly and clinically vulnerable, as they are more likely to suffer severe, and possibly fatal, respiratory illness. Nevertheless, anyone, regardless of age or medical background, is at risk of suffering a harsh illness.” Michali wishes to emphasise that **the best thing you can do to protect others from this disease is to take up the vaccine when you are offered it**. “Some of the vaccines on the market are boasting 95% efficacy rates. This means that receiving the vaccine dramatically reduces your risk of developing serious COVID-19 symptoms if you are exposed to the virus later down the line. Community transmission has been shown to be lower when severe COVID-19 symptoms do not present, so you are protecting your neighbours, parents, grandparents, and friends by receiving the vaccine.” Although experts are continuing to emphasise the importance of social distancing and wearing masks, these measures are not perfect, as the virus can still spread between people. The worry is not so much about individual cases, but rather, it is about reducing transmission in communities, as it is that type of transmission that will prevent us from ever seeing an end to this pandemic. Receiving the vaccine is the single most important step an individual can take to protect the community from the virus. Michali reflects in the concluding thoughts of the pamphlet, **“there is not much that we can control in times like these, but we need to take collective action to fight this pandemic! Taking up the vaccine when offered is the best action you can take for your family, friends, and for your community!”** Importantly, Michali wants people to remember that it is their responsibility to keep people in their community, especially those who are vulnerable to the disease, protected.

**Figure A.3.2:** Survey design : manipulation check at $T_1$.

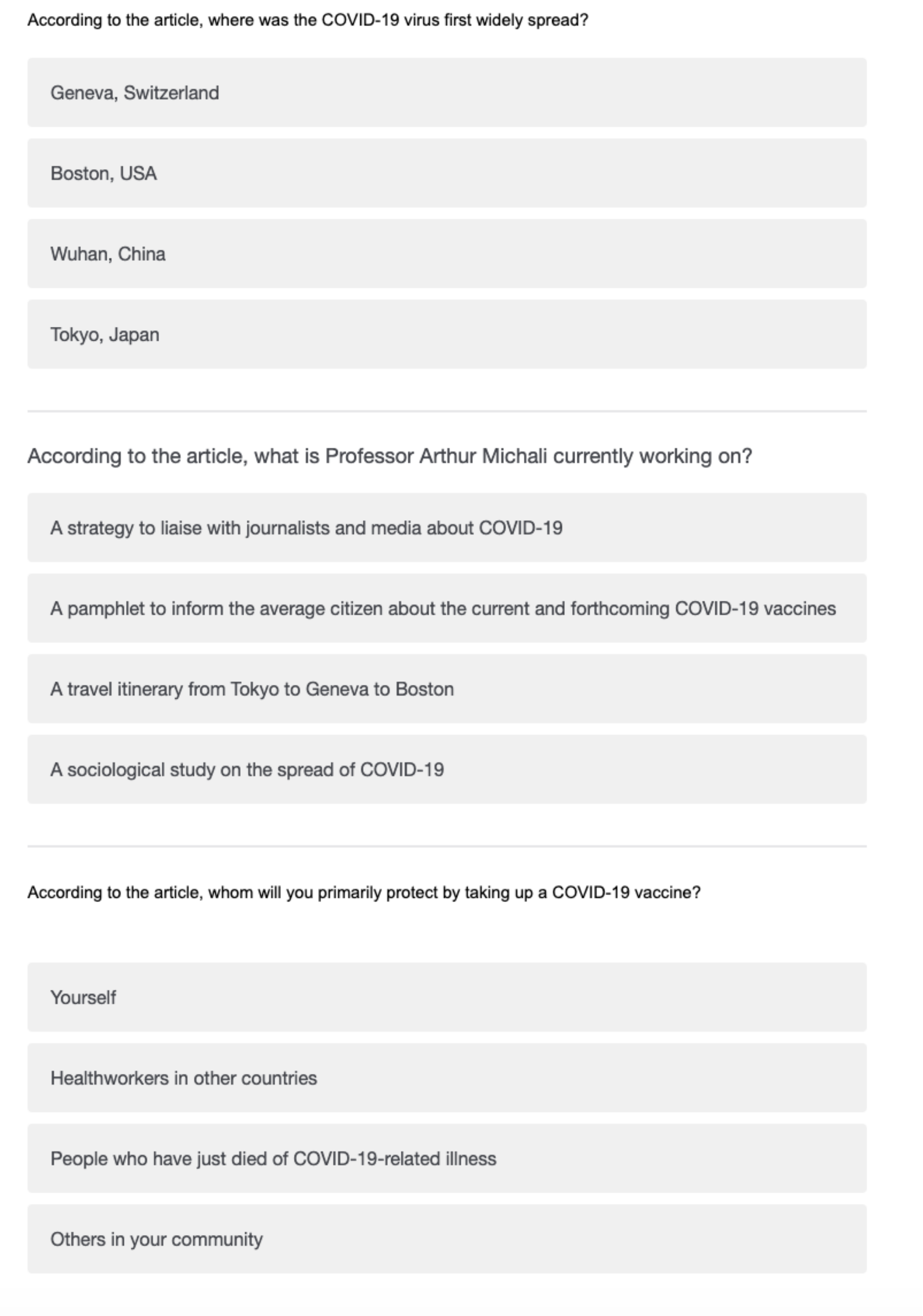

According to the article, where was the COVID-19 virus first widely spread?

- Geneva, Switzerland
- Boston, USA
- Wuhan, China
- Tokyo, Japan

According to the article, what is Professor Arthur Michali currently working on?

- A strategy to liaise with journalists and media about COVID-19
- A pamphlet to inform the average citizen about the current and forthcoming COVID-19 vaccines
- A travel itinerary from Tokyo to Geneva to Boston
- A sociological study on the spread of COVID-19

According to the article, whom will you primarily protect by taking up a COVID-19 vaccine?

- Yourself
- Healthworkers in other countries
- People who have just died of COVID-19-related illness
- Others in your community

**Figure A.3.3:** Survey design: demographic questions at $T_1$.

What is your ethnicity?

What is the highest level of education that you have completed?

In which region do you currently reside?

What is your yearly household income before tax?

Which political party do you consider yourself to be closest to?

Please indicate your attitudes to each of the following statements.

| | I disagree a lot | I disagree a little | I neither agree nor disagree | I agree a little | I agree a lot |
|---|---|---|---|---|---|
| In uncertain times, I usually expect the best. | O | O | O | O | O |
| I'm always optimistic about my future. | O | O | O | O | O |
| Overall, I expect more good things to happen to me than bad. | O | O | O | O | O |

Are you generally a person who tries to avoid taking risks or are you fully prepared to take risks?

Won't take risks — Ready to take risks

0 1 2 3 4 5 6 7 8 9 10

Have you been diagnosed with COVID-19 at any point?

How frequently do you follow government guidelines on face coverings when in shops?

- I never wear a face covering because I am exempt from wearing one.
- I never wear a face covering and I am not exempt from wearing one.
- Most of the time I do not wear a face covering.
- Half of the time I wear a face covering, half I do not.
- Most of the time I wear a face covering.
- I always wear a face covering.

How confident are you that face coverings are a useful tool in fighting the pandemic?

Not at all confident — Extremely confident

0 10 20 30 40 50 60 70 80 90 100

**Table A.2:** Factor analysis on the personal and general BJW factors, which produce the combined BJW factor.

Factor analysis/correlation

| Factor | Eigenvalue | Difference | Proportion | Cumulative |
|---|---|---|---|---|
| Factor1 | 0.79 | 1.04 | 1.46 | 1.46 |

Factor loadings (pattern matrix) and unique variances

| Variable | Factor1 | Uniqueness |
|---|---|---|
| General BJW | 0.63 | 0.61 |
| Personal BJW | 0.63 | 0.61 |

Scoring coefficients

| Variable | Factor1 |
|---|---|
| General BJW | 0.41 |
| Personal BJW | 0.41 |

Cronbach's alpha

| | |
|---|---|
| α | 0.69 |

**Table A.3:** Factor analysis on the skepticism outcomes at $T_0$.

Factor analysis/correlation

| Factor | Eigenvalue | Difference | Proportion | Cumulative |
|---|---|---|---|---|
| Factor1 | 2.57 | 2.60 | 1.10 | 1.10 |

Factor loadings (pattern matrix) and unique variances

| Variable | Factor1 | Uniqueness |
|---|---|---|
| Vaccine Useful | 0.86 | 0.26 |
| Accept Vaccine | 0.83 | 0.31 |
| Reduce Transmission | 0.64 | 0.59 |
| Prevent Illness | 0.85 | 0.28 |

Scoring coefficients

| Variable | Factor1 |
|---|---|
| Vaccine Useful | 0.35 |
| Accept Vaccine | 0.28 |
| Reduce Transmission | 0.12 |
| Prevent Illness | 0.31 |

| Cronbach's alpha | |
|---|---|
| α | 0.88 |

**Table A.4:** Factor analysis on the skepticism outcomes at $T_1$.

Factor analysis/correlation

| Factor | Eigenvalue | Difference | Proportion | Cumulative |
|---|---|---|---|---|
| Factor1 | 2.69 | 2.70 | 1.09 | 1.09 |

Factor loadings (pattern matrix) and unique variances

| Variable | Factor1 | Uniqueness |
|---|---|---|
| Vaccine Useful | 0.88 | 0.22 |
| Accept Vaccine | 0.82 | 0.33 |
| Reduce Transmission | 0.70 | 0.51 |
| Prevent Illness | 0.87 | 0.25 |

Scoring coefficients

| Variable | Factor1 |
|---|---|
| Vaccine Useful | 0.37 |
| Accept Vaccine | 0.23 |
| Reduce Transmission | 0.13 |
| Prevent Illness | 0.32 |

| Cronbach's alpha | |
|---|---|
| α | 0.89 |